\RequirePackage{fix-cm}
\documentclass[smallextended]{svjour3}       
\smartqed  
\usepackage{graphicx}
\usepackage{amssymb,amsfonts,amsmath}
\usepackage{Sweave}
\usepackage[round]{natbib}
\begin{document}

\title{Bayesian analysis for a class of beta mixed models
}


\author{Wagner Hugo Bonat \and Paulo Justiniano Ribeiro Jr \and Silvia Emiko Shimakura    
}


\institute{Universidade Federal do Paran\'a \at
           Dpto Estat\'istica-UFPR, CP 19.081, Curitiba, PR Brazil, 81.531-990 \\
              Tel.: +51-41-3361-3573\\
              Fax: +51-41-3361-3141\\
              \email{wbonat@ufpr.br}
}

\date{Received: date / Accepted: date}

\maketitle

\begin{abstract}
Generalized linear mixed models (GLMM) encompass large class of statistical models, 
with a vast range of applications areas. GLMM extends the linear mixed models allowing 
for different types of response variable. Three most common data types are continuous, counts and binary and standard distributions for these types of response variables are Gaussian, Poisson and Binomial, respectively. Despite that flexibility, there are situations where the response variable is continuous, but bounded, such as rates, percentages, indexes and proportions. In such situations the usual GLMM's are not adequate because bounds are ignored and 
the beta distribution can be used. Likelihood and Bayesian inference for beta mixed models are not 
straightforward demanding a computational overhead. 
Recently, a new algorithm for Bayesian inference called INLA (Integrated Nested Laplace Approximation) was proposed. 
INLA allows computation of many Bayesian GLMMs in a reasonable amount time allowing extensive comparison among models. 
We explore Bayesian inference for beta mixed models by INLA. We discuss the choice of prior distributions, sensitivity analysis and model selection measures through a real data set. The results obtained from INLA are compared with those obtained by an MCMC algorithm and likelihood analysis. 
We analyze data from an study on a life quality index of industry workers collected according to a hierarchical sampling scheme. Results show that the INLA approach is suitable and faster to fit the proposed beta mixed models 
producing results similar to alternative algorithms and with easier handling of modeling alternatives. 
Sensitivity analysis, measures of goodness of fit and model choice are discussed. 
\end{abstract}

\section{Introduction}
There has been an increased interest in the class of Generalized Linear Mixed Models (GLMM). One possible reason for such popularity is that GLMM combine Generalized Linear Models (GLM) \citep{Nelder1972} with Gaussian random effects, adding flexibility to the models and accommodating complex data structures such as hierarchical, repeated measures, longitudinal, among others which typically induce extra variability and/or dependence.
 
GLMMs can also be viewed as natural extension of Mixed Linear Models \citep{Pinheiro:2000}, allowing a wider class of probability distributions for response variables. Common choices are Gaussian for continuous data, Poisson and Negative Binomial for count data and Binomial for binary data. These three situations include the majority of applications within this class of models. Examples can be found in \citep{Breslow:1993} and \citep{Molenberghs:2005}.

Despite that flexibility, there are situations where the response variable is continuous and bounded above and below
such as rates, percentages, indexes and proportions. In such situations the traditional GLMM based on the Gaussian distribution, is not adequate, since bounding is ignored. An approach that has been used to model this type of data is based on the beta distribution. The beta distribution is very flexible with density function that can display quite different shapes, including left or right skewness, symmetric, J-shape, and inverted J-shape \citep{Da-Silva2011}.

Regression models for independent and identically distributed beta variable proposed by \cite{Paolino2001}, \cite{Kieschnick2003} and \cite{Ferrari2004}. The basic assumption is that the response follow a beta law whose expected value is related to a linear predictor through a link function, similarly to GLM's.
\citet{Cepeda2001}, \citet{Cepeda2005} and \citet{Simas2010} extend the model regressing both, the mean and the precision parameters with covariates. \citet{smithson:2006} explores beta regression with an application to IQ data. Methods for likelihood based inference and model assessment are proposed by \citet{Espinheira2008}, \citet{Espinheira2008a} and \citet{Rocha2010}. Bias corrections for likelihood estimators are developed by \citet{Ospina2006}, \citet{ospina_erratum:_2011} and \citet{Simas2010}. 
\citet{Branscum2007} adopts Bayesian inference to analyze virus genetic distances. Recently, \citet{Bonat2012} contrasted beta regression models with other approaches to model response variable on the unity interval, such that, Simplex, Kumaraswamy and Trans-Gaussian models. Results show there is no overall prominent model.

The beta regression is implemented by \texttt{betareg} package \citep{Cribari-neto2010} for the R environment for statistical computing \citep{R}. Extended functionality is added for bias correction, recursive partitioning and latent finite mixture \citep{grun2011}. Mixed and mixture models are further discussed by \citet{verkuilen_mixed_2011}.

For non independent data, development have been proposed in times series analysis by \citet{McKenzie1985}, \citet{Grunwald1993} and \citet{Rocha2008}. \citet{Da-Silva2011} use a Bayesian beta dynamic model for modeling and prediction of time series with an application to the Brazilian unemployment rates.
\citet{zuniga:2013} extend the beta model proposed by \citet{Ferrari2004} using a Bayesian approach. The authors considered two distributions for the random effects (Gaussian and t-Student) and several specifications for the prior distributions for parameters in the model.

\citet{Bonat2013} extend the beta model proposed by \citet{Ferrari2004} with the inclusion of Gaussian random effects, under a GLMM approach. Likelihood inference is based on two algorithms. The first uses the Laplace approximation to solve the integral in the likelihood function and the second uses an algorithm proposed by \citet{Lele2010} called data cloning. Authors analyzed two real data sets, with different structures for the random effects.
Likelihood inference under GLMM is non-trivial because of presence random effects and several procedures have been proposed. Approximate likelihood methods are adopted by \cite{Breslow:1993} and a Monte Carlo approach is adopted by  \cite{Chen2002}. Both come with a computational overhead. A popular approach is based upon a Bayesian framework using Markov Chain Monte Carlo (MCMC) algorithms with attempts to set non informative priors. \citet{zuniga:2013} perform Bayesian inference for beta mixed models using an MCMC algorithm. The Bayesian approach is attractive but requires specification of prior distributions, which is not straightforward, in particular for variance components.

Recently, \citet{Rue2009} introduced a novel numerical inference approach, the so-called Integrated Nested Laplace Approximation (INLA). INLA allows the computation of many Bayesian GLMMs in a reasonable amount of time, 
enabling for extensive comparisons of different models and prior distributions. 
\citet{Fong2010} used INLA for Bayesian analysis of several data sets and concluded that INLA is a very accurate algorithm and present results which help to guide the choice of prior distributions.

The main goal this paper is describe Bayesian inference for beta mixed models using INLA. 
We discuss the choice of prior distributions and measures of model comparisons. 
Results obtained from INLA are compared to those obtained using a Bayesian MCMC algorithm and a purely likelihood 
analysis. The modelling is illustrated through the analysis of a real dataset from a study on a life quality index of industry workers, with data collected according to a hierarchical sampling scheme. Additional care is given to choice of prior distributions for precision parameter of the beta law.

The structure this paper is the follows. In Section 2, we define the Bayesian beta mixed model, Section 3 we describe the Integrated Nested Laplace Approximation (INLA). In Section 4 the model is introduced for the motivating example and the results of the analyses are presented. We close with concluding remarks in Section 5.
\medskip

\section{Bayesian beta mixed model}
Bayesian beta mixed regression extends the beta regression model, as proposed by \citet{Ferrari2004}, by adding Gaussian distributed random effects to the linear predictor. 
Consider the response $Y_{ij}$ from group $i = 1, \ldots, N$ and replication $j = 1, \ldots, n_i$. $\mathbf{Y}_i$  is a $n_i-dimensional$ vector of measurements on the  $i^{th}$ group. Given a $q$-dimensional vector $\mathbf{b_i}$ of random effects distributed as $N(\mathbf{0}, Q(\boldsymbol{\tau}))$, the responses $Y_{ij}$ are conditionally independent with density function given by 
\begin{equation}
\pi_i(y_{ij}| \mathbf{b_i}, \mu_{ij}, \phi) =  \frac{\Gamma(\phi)}{ \Gamma(\mu_{ij} \phi) \Gamma( (1-\mu_{ij})\phi)} y_{ij}^{\mu_{ij} \phi - 1} (1 - y_{ij})^{(1-\mu_{ij})\phi -1}, \quad 0 < y < 1,
\end{equation}
where $0 < \mu < 1$ is the mean of the response variable and $\phi > 0$ is a dispersion parameter. 
Let $g(.)$ be a known link function with 
$g(\mu_{ij}) = \mathbf{x}_{ij}^T \boldsymbol{\beta} + \mathbf{z}_{ij}^T \mathbf{b_i}$, 
where $\mathbf{x}_{ij}$ and $\mathbf{z}_{ij}$ are vectors of covariates with dimensions $p$ and $q$, respectively, 
and $\boldsymbol{\beta}$ is a $p$-dimensional vector of unknown regression parameters. 
Assume that $\mathbf{b_i}|Q(\boldsymbol{\tau}) \sim N(\mathbf{0}, Q(\boldsymbol{\tau})^{-1})$, where the precision matrix 
$Q(\boldsymbol{\tau})$ depends on parameters $\boldsymbol{\tau}$. 
The model specification is completed assuming prior distributions for all parameters in the model, 
say $\mathbf{\theta} = (\boldsymbol{\beta}, \phi, \boldsymbol{\tau})$.

A flat improper prior is assumed for the intercept $\beta_0$. 
All other components of $\boldsymbol{\beta}$ are assumed to be independent $N(0,\sigma^2)$ with fixed precision 
$\sigma^{-2} = 0.0001$. For the parameters in the precision matrix $(\boldsymbol{\tau})$ 
we follow an approach adopted by \citet{Fong2010} based on \citet{Wakefield2009}. 
The basic idea is to specify a range for the more interpretable marginal distribution of $b_i$ 
and use this to derive the specification of the prior distributions. 
The approach is based on the result that if $b|\tau \sim N(0,\tau^{-1})$ and $\tau \sim Ga(a_1, a_2)$ then $b \sim t(0, a_2 / a_1, 2 a_1)$. To decide upon a prior, we define a range for a generic 
random effects $b$ and specify the degrees of freedom, $d$, and then solve for $a_1$ and $a_2$. The solution for a generic range, say $(-R,R)$, is $a_1 = d/2$ and $a_2 = R^2 d / (2 (t_{1-(1-q)/2}^d))^2$. 
In linear mixed effects model, $b$ is directly interpretable, while for beta models, it is more appropriate to think in terms of the marginal distribution of $exp(b)$. 
The prior distributions obtained this way are flat. 
For more detailed description see \citet{Fong2010} section $4.2$.

A flat $Ga(a_1 = 1,a_2 = 0.001)$ prior is choosen for $\phi$ 
as no result is known to aid its specification. 
The sensitivity to prior assumptions on the precision parameters of the beta distribution 
and of the random effects is a potentially a delicate issue under beta mixed models. 
\citet{zuniga:2013} considers several choices of prior distributions to $\phi$ 
but no sensitivity analysis is performed.
The idea here is to specify this Gamma distribution as the  
default choice and then to assess the sensitivity.

\section{Inference, model selection and sensitivity}
Bayesian inference on beta mixed models 
is not straightforward since the posterior distribution is not analytically available. 
Markov Chain Monte Carlo (MCMC) technique is the standard approach to fit such models \citep{zuniga:2013}. 
In practice, this approach comes with a wide range of problems in terms of convergence and computational time. 
Moreover, the implementation itself can be problematic, especially for end users who might not be experts in programming. 
Software platforms for fitting generic random effects models via MCMC, include JAGS 
\citep{Plummer03jags:a}, BayesX \citep{BayesX} and WinBUGS \citep{Lunn2000}, among others.

\citet{r-inla} is a newer tool for an end user 
based on the  INLA (Integrated Nested Laplace Approximation) approach  
for Bayesian inference on latent Gaussian models 
with focus on the posterior marginal distributions \citep{Rue2009}. 
INLA replaces MCMC simulations by accurate  
deterministic  approximations to posterior marginal distributions. 

A computational implementation called \texttt{inla}, available at \texttt{http://www.r-inla.org}, allows the user to conveniently perform approximate Bayesian inference in latent Gaussian models. The \textsl{R} package \texttt{INLA} serves as an interface to \texttt{inla} routines and its usage is similar to the \texttt{glm} function in \textsl{R} \citep{Roos2011}. 
Standard output provides marginal posterior densities for all parameters in the model and several measures of model goodness of fit.

The procedure of statistical analysis of a real data set, consists of specify the model, parameter estimation, comparisons among several models and 
evaluation results sensitivity, given the model specification. The second is tackled by INLA, 
and then the output includes several measures of model 
goodness of fit. The three more useful are, the Deviance Information Criterion (DIC), 
the log marginal likelihood (LML) and 
the conditional predictive ordinate (CPO), for more details, see \citet{Roos2011}. 

\citet{Roos2011} develop a general sensitivity measure based on the Hellinger distance to assess sensitivity of the posterior distributions with respect
to changes on the prior distributions for precision parameters. 
Such methods is adopted here to assess the 
sensitivity to the choice of the prior distribution for $\phi$ and for the precision of the random effects.
Following \citet{Roos2011}, for a default $\boldsymbol{\theta}_0$ and a shifted $\boldsymbol{\theta}$ prior value let
\begin{equation}
S(\boldsymbol{\theta}_0, \boldsymbol{\theta}) = \frac{H(post(\boldsymbol{\theta}_0), post(\boldsymbol{\theta}))}{H(pri(\boldsymbol{\theta}_0), pri(\boldsymbol{\theta}))} 
\end{equation}
denote the relative change on the posterior distribution with respect to changes on the prior distribution as measured 
by the Hellinger distance $H$, where $pri(\theta)$ is the prior distribution, $post(\theta)$ is the corresponding 
posterior distribution and 
\begin{equation*}
H(f,g) = \sqrt{ 1 - BC(f,g)}, \quad \text{where} \quad BC(f,g) = \int_{-\infty}^{\infty} \sqrt{f(u),g(u)}du.
\end{equation*}
The Hellinger distance is symmetric and measures the discrepancy between two densities $f$ and $g$.
It takes a maximal value of $1$ if $BC$ is equal to $0$ and
is equal to $0$ if and only if both densities are equal. 
The latter happens whenever the density $f$ assigns probability $0$ to every set to 
which the density $g$ assigns a positive  probability and vice versa.
For more detailed description see \citet{Roos2011}.

\section{Income and life quality of Brazilian industry workers}
The Brazilian industry sector \textit{worker's life quality index} (IQVT, acronym in Portuguese) is computed combining 25 indicators from eight thematic areas: housing, health, education, integral health and safety in the workplace, development of skills, value attributed to work, corporate social responsibility, stimulus to engagement and performance. The index is constructed following same premises as for the united nations human development index  \footnote{http://hdr.undp.org/en/humandev/}. The resulting values are in the unit interval and the closer to one the higher the worker's life quality in the industry.

A poll was conducted by Industry Social Service (\textit{Servi\c{c}o Social da Ind\'ustria - SESI}) in order to assess worker's life quality in the Brazilian industries. The survey included $365$ companies on eight Brazilian federative units among the total of 26 states plus the Federal District. The data analysis considers two covariates related to the companies for which the impact on IQVT is of particular interest, namely, company average \textit{income} and \textit{size}. The first is given by the total of salaries divided by the number of workers expressing the capacity to fulfill individual basic needs such as food, health, housing and education. The second can be indirectly related to the capability of managing and providing quality of life.

The relevant question for the study and main goal here is to specify a suitable model to assess the influence of these two covariates on the IQVT. The federative unit where the company based is expected to influence the index considering varying local legislations, taxing and further economic and political conditions. This is accounted by including a random effect, regarding the eight states as a sample of the federative units.

\begin{figure}[htbp]
\centering
\includegraphics{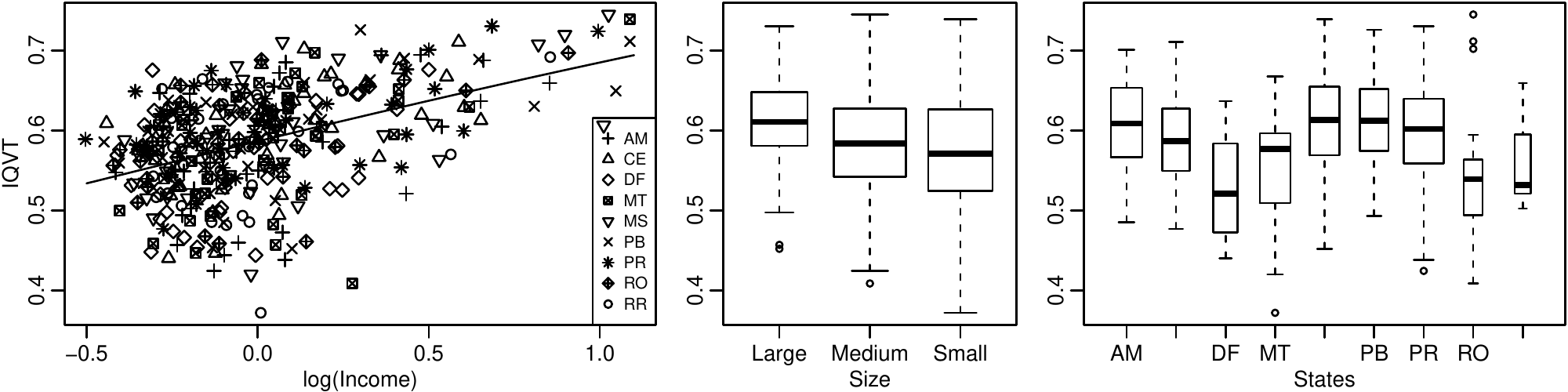}
\caption{Scatterplot and boxplot relating IQVT with (centred) log income, company size and state.}
\label{fig:descritivaIQVT}
\end{figure}

Relations between the IQVT and the covariates income, size and with the states included 
in the survey are shown on Figure~\ref{fig:descritivaIQVT} which suggests all are potentially relevant. 
The income is expressed in logarithmic scale centered around the average.

\setkeys{Gin}{width=0.99\textwidth}

The Bayesian beta random effects model for IQVT is given by
\begin{align*}
\label{modelo}
  Y_{ij} | \mathbf{b}_i &\sim Beta(\mu_{ij} , \phi) \\
  g(\mu_{ij}) &= (\beta_0 + b_{i1}) + \beta_1 Medium_{ij} + \beta_2 Small_{ij} + (\beta_3 + b_{i2}) Income_{ij} \\
  \mathbf{b}_i &\sim NMV(\mathbf{0}, \Sigma)  \mbox{  with  } \Sigma = \left[\begin{array}{rr}
1/ \tau_1^2 &\rho\\
\rho& 1 / \tau_2^2 \\
\end{array}\right]
\end{align*}
parametrize with $\beta_0$ associated with large size companies with differences $\beta_1$ and $\beta_2$ to the medium and small size, respectively. Random effects include an intercept $b_{i1}$ and a slope $b_{i2}$ associated with the covariate \textit{income}. The vector of model parameters are the regression coefficients $(\beta_0, \beta_1, \beta_2, \beta_3)$, the random effects covariance parameters $(\tau_1^2, \tau_2^2, \rho)$ and dispersion parameter $\phi$ from beta law. The logit  $g(\mu_{ij}) = \log\{\mu_{ij}/(1 -\mu_{ij})\}$ link function is used.
The specification the Bayesian beta mixed model is completed by specifying the prior distributions for the model parameters. Following the remarks at Section~$2$  a flat improper prior is assumed $\beta_0$. All other components of $\beta$ are assumed to be independent zero-mean $N(0,\sigma^2)$ with fixed precision $\sigma^{-2} = 0.0001$. For the parameter $\phi$ we assumed a flat $Ga(a_1 = 1, a_2 = 0.0001)$ distribution. For the parameters indexing the random effects $\Sigma = Q^{-1}$, we assumed that $ Q \sim W_q(r,S)$, where $W_q(r,S)$ denote the Wishart distribution, $r$ and $S$ to be chosen as in the univariate case. Specifically, we assumed that $r = 5$ and a diagonal $S$ 
with elements $0.001487$ and $0.005$, reducing to a $Ga(a_1 = 0.5, a_2 = 0.001487)$ 
when fitting the random intercept model.

A sequence of sub-models are defined in order to assess the effects of interest. Model~1 is a null model with just the intercept coefficent. 
Model~2 and~3 adds the covariates \textit{size} and \textit{income}, in this order. 
Model~4 and~5 adds random effects related to the \textit{States} to the intercept and the \textit{income} coefficient, respectively. 
The latter is the largest model considered here. 
A sequence of nested  models are defined for comparison and detection of the relevant effects.
Large size companies are considered as the baseline for the categorical covariate \textit{size}.
Table~\ref{tab:comparativoIQVT} shows the posterior means for the model parameters 
and model fitting measures given by  the deviance information criterion (DIC), log marginal likelihood (LML) and conditional predictive ordinate (CPO),
all obtained with \texttt{INLA}. 


 \begin{table}[] 
 \centering 
 \caption{Posterior means, LML, DIC and CPO for the fitted models.}  
 \label{tab:comparativoIQVT} 
 \begin{tabular}{lccccc} 
   \hline 
 Parameter & Model 1 & Model 2 & Model 3 & Model 4 & Model 5 \\ 
   \hline 
 $\beta_0$ & 0.35 & 0.45 & 0.43 & 0.40 & 0.40 \\  
   $\beta_1$ &  & -0.11 & -0.09 & -0.07 & -0.07 \\  
   $\beta_2$ &  & -0.16 & -0.14 & -0.13 & -0.14 \\  
   $\beta_3$ &  &  & 0.42 & 0.47 & 0.46 \\  
   $\phi$ & 53.92 & 56.44 & 72.16 & 93.37 & 93.28 \\  
   $\tau_1^2$ &  &  &  & 63.65 & 90.33 \\  
   $\tau_2^2$ &  &  &  &  & 532.73 \\  
   $\rho$ &  &  &  &  & 0.75 \\  
 \hline \multicolumn{6}{c}{Goodness-of-fit} \\ 
 LML & 466.02 & 461.57 & 500.11 & 534.40 & 359.82 \\  
   DIC & -941.11 & -955.72 & -1044.58 & -1130.79 & -1129.29 \\  
   CPO & -1.29 & -1.31 & -1.43 & -1.55 & -1.55 \\  
    \hline 
 \end{tabular} 
 \end{table} 
 
 Results for models 1-3 confirm the relevance of the covariates. The increasing values for 
 average posterior of $\phi$, from $53.92$ on model~1 to $72.16$ on model~3, confirms further explanation of the data variability by the covariates.
The random intercept clearly improves the model fit, capturing the variability of the IQVT among the states. 
The addition of random slope did not prove relevant. 
All model fitting measures favors model~4 for which we report further analysis. 

\begin{figure}[htbp]
\centering
\includegraphics{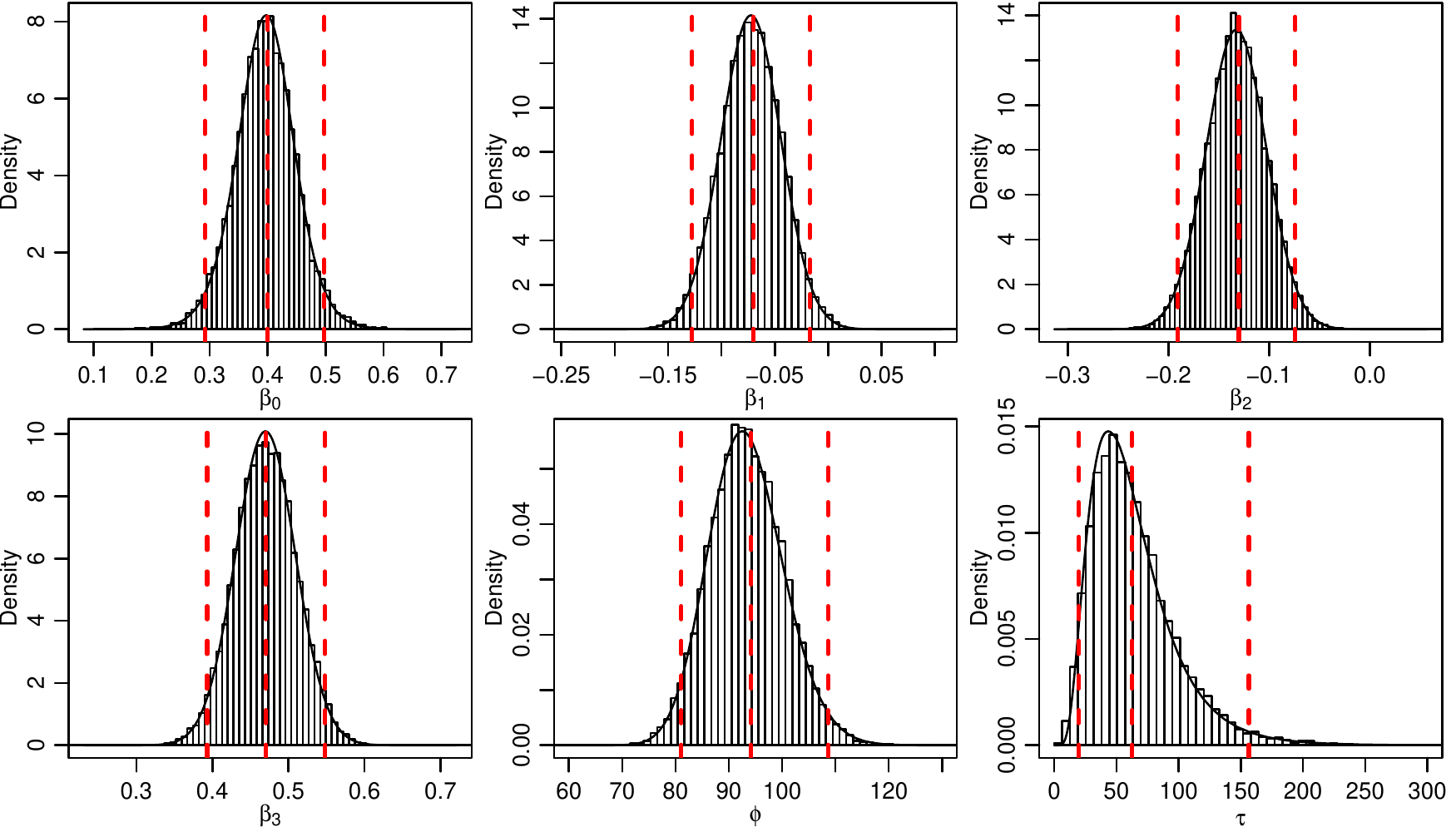}
\caption{Marginal posterior densities from INLA and MCMC outputs. The vertical dashed lines correspond to point estimates and profile likelihood intervals.}
\label{fig:posterioriIQVT}
\end{figure}

Figure~\ref{fig:posterioriIQVT} shows posterior distributions from INLA and a MCMC output from JAGS running three chains of 500,000 samples
 with a burn-in of 10,000 interactions and saving one of each 100 simulations.
We also compared INLA results with likelihood point estimates and profile intervals.
Figure~2 suggests that all approaches produced similar results.
This is also assured by the results in Table~\ref{tab:coverageIQVT} where
the second and third columns provide the proportion of MCMC samples which
falls into the profile likelihood interval and credibility intervals from INLA, respectively.
The last column is the probability between the limits of the profile likelihood interval
computed on the INLA marginal distribution.
These results indicates the flat prior has little impact on the respective posterior distribution. 
The INLA and MCMC algorithms are similar in the inferential purposes but INLA is much faster and easier to use than MCMC. 

\setkeys{Gin}{width=0.95\textwidth}

\begin{table}[]
\centering
\caption{Comparison on the intervals obtained by different methods.} 
\label{tab:coverageIQVT}
\begin{tabular}{cccc}
  \hline
Parameter & MCMC.in.Profile & MCMC.in.INLA & Profile.in.INLA \\ 
  \hline
$\beta_0$ & 0.9415 & 0.9435 & 0.9487 \\ 
  $\beta_1$ & 0.9457 & 0.9471 & 0.9490 \\ 
  $\beta_2$ & 0.9496 & 0.9509 & 0.9491 \\ 
  $\beta_3$ & 0.9465 & 0.9471 & 0.9491 \\ 
  $\phi$ & 0.9491 & 0.9510 & 0.9480 \\ 
  $\tau_1^2$ & 0.9446 & 0.9433 & 0.9511 \\ 
   \hline
\end{tabular}
\end{table}

\begin{figure}[htbp]
\centering
\includegraphics{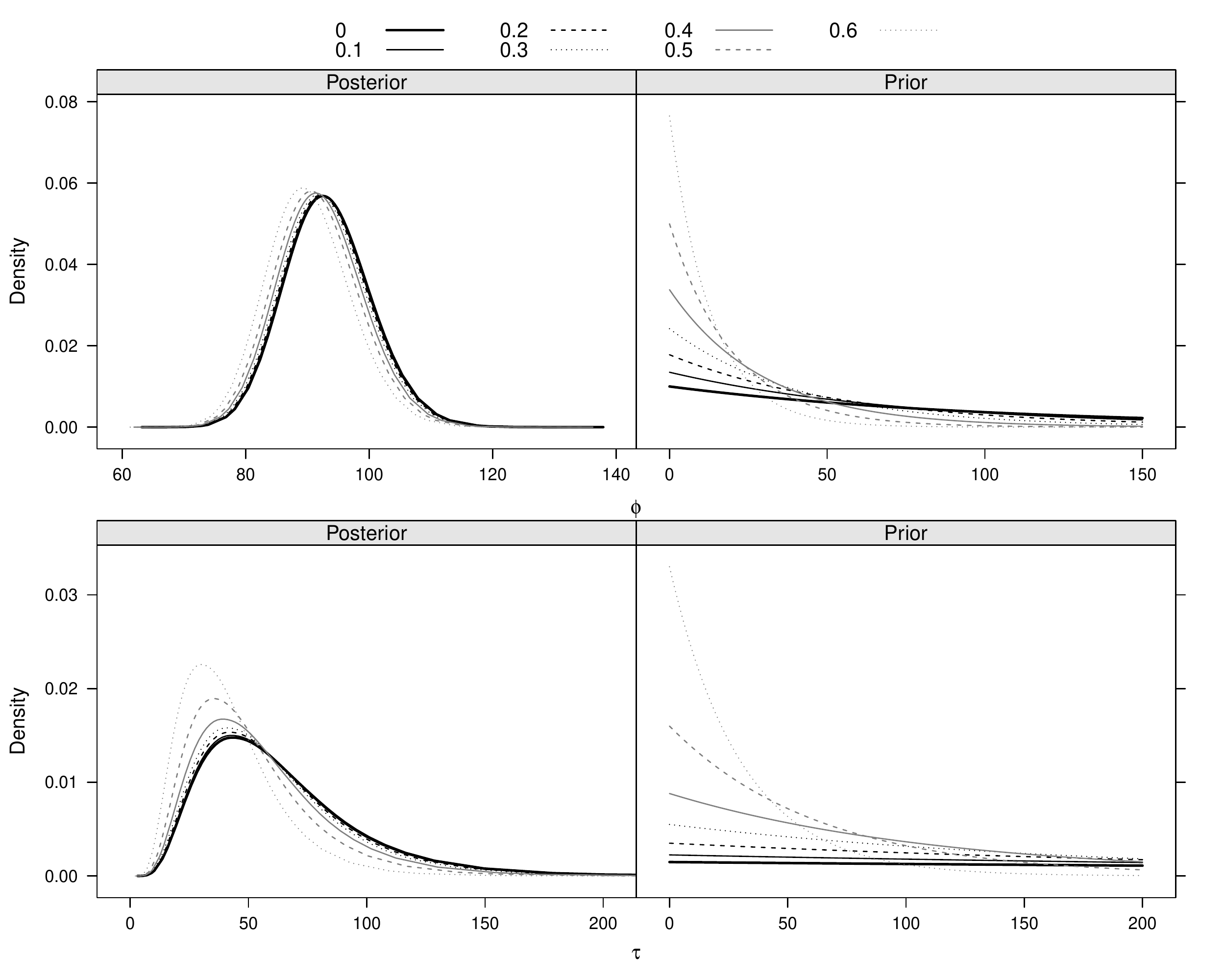}
\caption{Posterior (left) and prior (right) distributions with Hellinger distances to the default prior ranging from 0.1 to 0.6 for the parameters $\phi$ (top) and $\tau$ (bottom).}
\label{fig:post}
\end{figure}

We conclude the analysis assessing sensitivity to the choice the prior distributions.
Following \citet{Roos2011}, we investigate the sensitivity by measuring the Hellinger distance and focusing 
only on the parameters $\phi$ and $\tau$ since the choice of prior is standard
for the regression coefficients $\beta's$. 
To assess sensitivity we choose a set of prior distributions 
with determined Hellinger distance from the default prior, refit the model 
under those priors and compute the  Hellinger distances between the corresponding posterior distributions. 
For example, by choosing $\phi \sim Ga(b1 = 1, b2 = 0.0135)$ the Hellinger distance from the 
 default $\phi_0 \sim Ga(a1 = 1, a2 = 0.001)$ prior
is $H(Ga(a_1 = 1, a_2 = 0.001), Ga(b_1 = 1, b_2 = 0.0135)) = 0.1058$
whereas the Hellinger distance between the posteriors is $0.0100$ 
and $S(\phi_0, \phi) = 0.0945$, i.e, the distance between the posterior distributions
  is only about one tenth of the distance between the prior distributions reflecting a major 
  effect of the that and a little impact of choosing either prior. 
Table \ref{tab:div} shows the hyperparameters obtained 
for priors with Hellinger distances from  about $0.1$ to $0.6$ 
and the corresponding Hellinger distances between priors, posteriors
and $S(\cdot,\cdot)$. The distributions are plotted in 
Figure~\ref{fig:post}.

\begin{table}[]
\centering
\caption{Hellinger distances between the prior and posterior distributions from the ones obtained with the default prior.} 
\label{tab:div}
\begin{tabular}{cccc}
  \hline
Priori & HL Prior & HL Post & $S(post,pri)$ \\ 
  \hline
$ \phi \sim Ga(b_1 = 1, b_2 = 0.0135)$ & 0.1058 & 0.0100 & 0.0945 \\ 
  $ \phi \sim Ga(b_1 = 1, b_2 = 0.0178)$ & 0.2005 & 0.0200 & 0.0998 \\ 
  $ \phi \sim Ga(b_1 = 1, b_2 = 0.0242)$ & 0.3005 & 0.0346 & 0.1153 \\ 
  $ \phi \sim Ga(b_1 = 1, b_2 = 0.0338)$ & 0.4006 & 0.0583 & 0.1455 \\ 
  $ \phi \sim Ga(b_1 = 1, b_2 = 0.050)$ & 0.5046 & 0.0975 & 0.1932 \\ 
  $ \phi \sim Ga(b_1 = 1, b_2 = 0.0765)$ & 0.6004 & 0.1628 & 0.2711 \\ 
  \hline
  $ \tau \sim Ga(b_1 = 0.5, b_2 = 0.00225)$ & 0.1030 & 0.0100 & 0.0971 \\ 
  $ \tau \sim Ga(b_1 = 1, b_2 = 0.0035)$ & 0.2086 & 0.0245 & 0.1174 \\ 
  $ \tau \sim Ga(b_1 = 1, b_2 = 0.0055)$ & 0.3085 & 0.0458 & 0.1485 \\ 
  $ \tau \sim Ga(b_1 = 1, b_2 = 0.0088)$ & 0.4017 & 0.0812 & 0.2022 \\ 
  $ \tau \sim Ga(b_1 = 1, b_2 = 0.016)$ & 0.5031 & 0.1543 & 0.3066 \\ 
  $ \tau \sim Ga(b_1 = 1, b_2 = 0.033)$ & 0.6022 & 0.2827 & 0.4694 \\ 
   \hline
\end{tabular}
\end{table}

The results show that the models are more sensitive to the choice of prior for the parameter $\tau$. 
For the parameter $\phi$ even with the rather large distance of 0.6 between prior distributions
 the corresponding distance between the posterior distributions is substantially reduced to $0.1628$. 
 The same distance between priors for the parameter $\tau$ 
 still reduces to $0.2827$. 
The posterior distributions in Figure~\ref{fig:post} are similar for all prior distributions. 
Comparatively, the parameter $\tau$ is  more sensitive to the choice of prior distribution,
 however still with similar posterior distributions even with large difference between priors.

\setkeys{Gin}{width=0.9\textwidth}

Table \ref{tab:compara} compare summary results 
of models with default prior default and with the largest Hellinger distance from the  
default prior.
For $\phi$ the posterior mean changed from  $93.37$ to $90.18$, a difference is only $3.53\%$
whereas for $\tau$ they change from $63.65$ to $41.86$ with a difference of $52.05\%$.
Despite such differences, the practical conclusions on effects of relevance are unchanged
since the changes are very small for the regression parameters.
The relevance of the random effects in the model remains important.


\begin{table}[]
\centering
\caption{Summaries of the fitted models with Hellinger distances of 0.6 from the default prior for parameters $\phi$ and $\tau$.} 
\label{tab:compara}
\begin{tabular}{lcccccc}
  \hline
 & Default & Std. Err & $ H(\phi_D,\phi_{0.6}) = 0.6 $ & Std. Err & $ H(\tau_D,\tau_{0.6}) = 0.6$ & Std. Err \\ 
  \hline
$\beta_0$ & 0.3965 & 0.0520 & 0.3966 & 0.0518 & 0.3950 & 0.0607 \\ 
  $\beta_1$ & -0.0724 & 0.0282 & -0.0725 & 0.0287 & -0.0719 & 0.0282 \\ 
  $\beta_2$ & -0.1327 & 0.0299 & -0.1327 & 0.0304 & -0.1320 & 0.0299 \\ 
  $\beta_3$ & 0.4700 & 0.0396 & 0.4695 & 0.0403 & 0.4718 & 0.0396 \\ 
  $\phi$ & 93.3746 & 7.0019 & 90.1881 & 6.7830 & 93.5137 & 6.9848 \\ 
  $\tau_1^2$ & 63.6521 & 34.2202 & 64.8726 & 34.3975 & 41.8625 & 20.8770 \\ 
   \hline
\end{tabular}
\end{table}




\section{Conclusion}
This paper reports results of a Bayesian analysis of beta mixed models 
comparing results obtained with the INLA method with the ones obtained with an MCMC
algorithm and purely likelihood analysis. 
Emphasis is placed on the specification and sensitivity of priors for the Beta dispersion parameter and
the precision of the random effects.

Results of the analysis of the index of life quality for the worker's on the Brazilian industrial sector 
indicates company size and average income are both relevant for the quality of life, as well as the effect of the states
captured by adding a random intercept to the regression model.
The analysis consisted of fitting several models with one final model chosen according to three 
criteria of model comparisons -- LML, DIC and CPO. 
All criteria points to the same model choice.
Summary results obtained with INLA are similar with the ones obtained with 
MCMC and likelihood analysis showing the substantial gain in the computational burden 
makes INLA an attractive choice for inference which allowing for several modeling alternatives 
to be investigated.

The sensitivity analysis was conducted for 
the dispersion parameters in the Bayesian beta mixed model using the Hellinger divergence
as a measure of the distance between prior and posterior distributions. 
Our results show that the Beta dispersion parameter $\phi$ is insensitive to the  choice of prior. 
Slightly more sensitive is the parameter $\tau$ related to the random effects, 
but the overall results and conclusions remains unchanged for the alternative priors.



\begin{acknowledgements}
Milton Matos de Souza and Sonia Beraldi de Magalh\~{a}es  from 
\textit{Servi\c{c}o Social da Ind\'ustria (SESI)} for the IQVT data.
\end{acknowledgements}

\bibliographystyle{dcu}
\bibliography{BonatBayes}   

\end{document}